# A Bayesian Neural Network Approach to identify Stars and AGNs observed by XMM Newton ★

Sarvesh Gharat,[1]† and Bhaskar Bose[2]
[1] *Centre for Machine Intelligence and Data Science, Indian Institute of Technology Bombay, 400076, Mumbai, India*
[2] *Smart Mobility Group, Tata Consultancy Services, 560067, Bangalore, India*

**ABSTRACT**
In today's era, a tremendous amount of data is generated by different observatories and manual classification of data is something which is practically impossible. Hence, to classify and categorize the objects there are multiple machine and deep learning techniques used. However, these predictions are overconfident and won't be able to identify if the data actually belongs to the trained class. To solve this major problem of overconfidence, in this study we propose a novel Bayesian Neural Network which randomly samples weights from a distribution as opposed to the fixed weight vector considered in the frequentist approach. The study involves the classification of Stars and AGNs observed by XMM Newton. However, for testing purposes, we consider CV, Pulsars, ULX, and LMX along with Stars and AGNs which the algorithm refuses to predict with higher accuracy as opposed to the frequentist approaches wherein these objects are predicted as either Stars or AGNs. The proposed algorithm is one of the first instances wherein the use of Bayesian Neural Networks is done in observational astronomy. Additionally, we also make our algorithm to identify stars and AGNs in the whole XMM-Newton DR11 catalogue. The algorithm almost identifies 62807 data points as AGNs and 88107 data points as Stars with enough confidence. In all other cases, the algorithm refuses to make predictions due to high uncertainty and hence reduces the error rate.

**Key words:** methods: data analysis – methods: observational – methods: miscellaneous

## 1 INTRODUCTION

Since the last few decades, a large amount of data is regularly generated by different observatories and surveys. The classification of this enormous amount of data by professional astronomers is time-consuming as well as practically impossible. To make the process simpler, various citizen science projects (Desjardins et al. 2021) (Cobb 2021) (Allf et al. 2022) (Faherty et al. 2021) are introduced which has been reducing the required time by some extent. However, there are many instances wherein classifying the objects won't be simple and may require domain expertise.

In this modern era, wherein Machine Learning and Neural Networks are widely used in multiple fields, there has been significant development in the use of these algorithms in Astronomy. Though these algorithms are accurate with their predictions there is certainly some overconfidence (Kristiadi et al. 2020) (Kristiadi et al. 2021) associated with it. Besides that, these algorithms tend to classify every input as one of the trained classes (Beaumont & Haziza 2022) irrespective of whether it actually belongs to those trained classes eg: The algorithm trained to classify stars will also predict AGNs as one of the stars. To solve this major issue, in this study we propose a Bayesian Neural Network (Jospin et al. 2022) (Charnock et al. 2022) which refuses to make a prediction whenever it isn't confident about its predictions. The proposed algorithm is implemented on the data collected by XMM-Newton (Jansen et al. 2001). We do a binary classification to classify Stars and AGNs (Małek et al. 2013) (Golob et al. 2021). Additionally to test our algorithm with the inputs which don't belong to the trained class we consider data observed from CV, Pulsars, ULX, and LMX. Although, the algorithm doesn't refuse to predict all these objects, but the number of objects it predicts for these 4 classes is way smaller than that of trained classes.

For the trained classes, the algorithm gives its predictions for almost 64% of the data points and avoids predicting the output whenever it is not confident about its predictions. The achieved accuracy in this binary classification task whenever the algorithm gives its prediction is 98.41%. On the other hand, only 14.6% of the incorrect data points are predicted as one of the classes by the algorithm. The percentage decrease from 100% to 14.6% in the case of different inputs is what dominates our model over other frequentist algorithms.

## 2 METHODOLOGY

In this section, we discuss the methodology used to perform this study. This section is divided into the following subsections.

- Data Collection and Feature Extraction
- Model Architecture
- Training and Testing

---

★ Based on observations obtained with XMM-Newton, an ESA science mission with instruments and contributions directly funded by ESA Member States and NASA
† E-mail: sarveshgharat19@gmail.com

| Class | Catalogue |
|---|---|
| AGN | VERONCAT (Véron-Cetty & Véron 2010) |
| LMX | NGC3115CXO (Lin et al. 2015) |
|  | RITTERLMXB (Ritter & Kolb 2003) |
|  | LMXBCAT (Liu et al. 2007) |
|  | INTREFCAT (Ebisawa et al. 2003) |
|  | M31XMMXRAY (Stiele et al. 2008) |
|  | M31CFCXO (Hofmann et al. 2013) |
|  | RASS2MASS (Haakonsen & Rutledge 2009) |
| Pulsars | ATNF (Manchester et al. 2005) |
|  | FERMIL2PSR (Abdo et al. 2013) |
| CV | CVC (Drake et al. 2014) |
| ULX | XSEG (Drake et al. 2014) |
| Stars | CSSC (Skiff 2014) |

**Table 1.** Catalogues used to create labeled data

| Class | Training Data | Test Data |
|---|---|---|
| AGN | 8295 | 2040 |
| LMX | 0 | 49 |
| Pulsars | 0 | 174 |
| CV | 0 | 36 |
| ULX | 0 | 261 |
| Stars | 6649 | 1628 |
| Total | 14944 | 4188 |

**Table 2.** Data distribution after cross-matching all the data points with catalogs mentioned in Table 1

## 2.1 Data Collection and Feature Extraction

In this study, we make use of data provided in "XMM-DR11 SEDs" Webb et al. (2020). We further cross-match the collected data with different vizier (Ochsenbein et al. 2000) catalogs. Please refer to Table 1 to view all the catalogs used in this study. As the proposed algorithm is a "supervised Bayesian algorithm", this happens to be one of the important steps for our algorithm to work.

The provided data has 336 different features that can increase computational complexity by a larger extent and also has a lot of missing data points. Therefore in this study, we consider a set of 18 features corresponding to the observed source. The considered features for all the sources are available on our Github repository, more information of which is available on the official webpage [1] of the observatory. After cross-matching and reducing the number of features, we were left with a total of 19136 data points. The data distribution can be seen in Table 2. We further also plot the sources (Refer Figure1) based on their "Ra" and "Dec" to confirm if the data coverage of the considered sources matches with the actual data covered by the telescope.

---

[1] http://xmmssc.irap.omp.eu/Catalogue/4XMM-DR11/col_unsrc.html

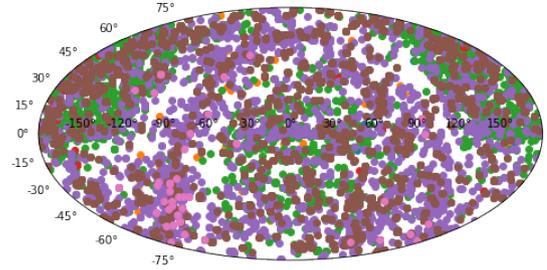

**Figure 1.** Sky map coverage of considered data points

The collected data is further classified into train and test according to the 80 : 20 splitting condition. The exact number of data points is mentioned in Table 2

## 2.2 Model Architecture

The proposed model has 1 input, hidden and output layers (refer Figure 2) with 18, 512, and 2 neurons respectively. The reason for having 18 neurons in the input layer is the number of input features considered in this study. Further, to increase the non-linearity of the output, we make use of "Relu" (Fukushima 1975) (Agarap 2018) as an activation function for the first 2 layers. On the other hand, the output layer makes use of "Softmax" to make the predictions. This is done so that the output of the model will be the probability of image belonging to a particular class (Nwankpa et al. 2018) (Feng & Lu 2019).

The "optimizer" and "loss" used in this study are "Adam" (Kingma et al. 2020) and "Trace Elbo" (Wingate & Weber 2013) (Ranganath et al. 2014) respectively. The overall idea of BNN (Izmailov et al. 2021) (Jospin et al. 2022) (Goan & Fookes 2020) is to have a posterior distribution corresponding to all weights and biases such that, the output distribution produced by these posterior distributions is similar to that of the categorical distributions defined in the training dataset. Hence, convergence, in this case, can be achieved by minimizing the KL divergence between the output and the categorical distribution or just by maximizing the ELBO (Wingate & Weber 2013) (Ranganath et al. 2014). We make use of normal distributions which are initialized with random mean and variance as prior (Fortuin et al. 2021), along with the likelihood derived from the data to construct the posterior distribution.

## 2.3 Training and Testing

The proposed model is constructed using Pytorch (Paszke et al. 2019) and Pyro (Bingham et al. 2019). The training of the model is conducted on Google Colaboratory, making use of NVIDIA K80 GPU (Carneiro et al. 2018). The model is trained over 2500 epochs with a learning rate of 0.01. Both these parameters i.e number of epochs and learning rate has to be tuned and are done by iterating the algorithm multiple times with varying parameter values.

The algorithm is further asked to make 100 predictions corresponding to every sample in the test set. Every time it makes the prediction, the corresponding prediction probability varies. This is due to random sampling of weights and biases from the trained distributions. Further, the algorithm considers the "mean" and "standard deviation" corresponding to those probabilities to make a decision as to proceed with classification or not.

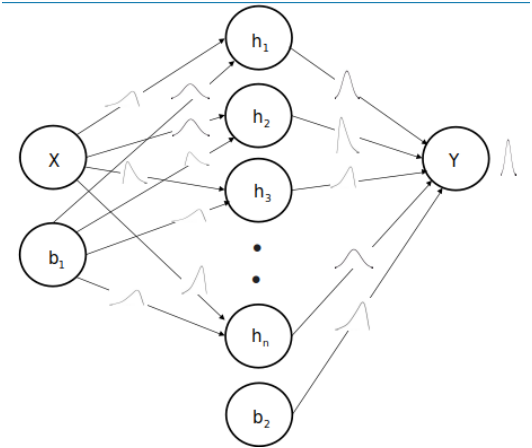

**Figure 2.** Model Architecture

|  | AGN | Stars |
|---|---|---|
| AGN | 1312 | 6 |
| Stars | 31 | 986 |

**Table 3.** Confusion Matrix for classified data points

| Class | Precision | Recall | F1 Score |
|---|---|---|---|
| AGN | 0.99 | 0.97 | 0.98 |
| Stars | 0.97 | 0.99 | 0.98 |
| Average | 0.98 | 0.98 | 0.98 |

**Table 4.** Classification report for classified data points

## 3 RESULTS AND DISCUSSION

The proposed algorithm is one of the initial attempts to implement "Bayesian Neural Networks" in observational astronomy which has shown significant results. The algorithm gives the predictions with an accuracy of more than 98% whenever it agrees to make predictions for trained classes.

Table 3 represents confusion matrix of classified data. To calculate accuracy, we make use of the given formula.

$$\text{Accuracy} = \frac{a_{11} + a_{22}}{a_{11} + a_{12} + a_{21} + a_{22}} \times 100$$

In our case, the calculated accuracy is

$$\text{Accuracy} = \frac{1312 + 986}{1312 + 6 + 31 + 986} \times 100 = 98.4\%$$

As accuracy is not the only measure to evaluate any classification model, we further calculate precision, recall and f1 score corresponding to both the classes as shown in Table 4

Although, the obtained results from simpler "BNN" can be obtained via complex frequentist models, the uniqueness of the algorithm is that it agrees to classify only 14% of the unknown classes as one of the trained classes as opposed to frequentist approaches wherein all those samples are classified as one of these classes. Table 5 shows the percentage of data from untrained classes

| Class | AGN | Star |
|---|---|---|
| CV | 13.8 % | 0 % |
| Pulsars | 2.3 % | 6.3 % |
| ULX | 14.9 % | 6.5 % |
| LMX | 2 % | 26.5 % |
| Total | 9.4 % | 7.8 % |

**Table 5.** Percentage of misidentified data points

which are predicted as a Star or a AGN.

As the algorithm gives significant results on labelled data, we make use of it to identify the possible Stars and AGNs in the raw data [2]. The algorithm almost identifies almost 7.1% of data as AGNs and 10.04% of data as AGNs. Numerically, the number happens to be 62807 and 88107 respectively. Although, there's high probability that there exists more Stars and AGNs as compared to the given number the algorithm simply refuses to give the prediction as it isn't enough confident with the same.

## 4 CONCLUSIONS

In this study, we propose a Bayesian approach to identify Stars and AGNs observed by XMM Newton. The proposed algorithm avoids making predictions whenever it is unsure about the predictions. Implementing such algorithms will help in reducing the number of wrong predictions which is one of the major drawbacks of algorithms making use of the frequentist approach. This is an important thing to consider as there always exists a situation wherein the algorithm receives an input on which it is never trained. The proposed algorithm also identifies 62807 Stars and 88107 AGNs in the data release 11 by XMM-Newton.

## 5 CONFLICT OF INTEREST

The authors declare that they have no conflict of interest.

## DATA AVAILABILITY

The raw data used in this study is publicly made available by XMM Newton data archive. All the codes corresponding to the algorithm and the predicted objects along with the predictions will be publicly made available on "Github" and "paperswithcode" by June 2023.

---

[2] http://xmmssc.irap.omp.eu/Catalogue/4XMM-DR11/col_unsrc.html

This paper has been typeset from a T{_E}X/L{^A}T{_E}X file prepared by the author.